\documentclass[aps,prc,groupedaddress,showpacs,manuscript]{revtex4-1}
\usepackage{graphicx}

\usepackage{colordvi}

\begin{document}

\title{Rapidity distribution of protons from the potential version of UrQMD model and the traditional coalescence afterburner}

\author {Qingfeng Li$\, ^{1}$\footnote{E-mail address: liqf@hutc.zj.cn},
Yongjia Wang$\, ^{1}$,
Xiaobao Wang$\, ^{1}$,
and Caiwan Shen$\, ^{1}$,
}

\affiliation{
1) School of Science, Huzhou University, Huzhou 313000, P.R. China \\
 }
\date{\today}

\begin{abstract}

Rapidity distributions of both E895 proton data at AGS energies and NA49 net proton data at SPS energies can be described reasonably well with a potential version of the UrQMD in which mean-field potentials for both pre-formed hadrons and confined baryons are considered, with the help of a traditional coalescence afterburner in which one parameter set for both relative distance $R_0$ and relative momentum $P_0$, (3.8 fm, 0.3 GeV$/$c), is used. Because of the large cancellation between the expansion in $R_0$ and the shrinkage in $P_0$ through the Lorentz transformation, the relativistic effect in clusters has little effect on the rapidity distribution of free (net) protons. Using a Woods-Saxon-like function instead of a pure logarithmic function as seen by FOPI collaboration at SIS energies, one can fit well both the data at SIS energies and the UrQMD calculation results at AGS and SPS energies. Further, it is found that for central Au+Au or Pb+Pb collisions at top SIS, SPS and RHIC energies, the proton fractions in clusters are about 33$\%$, 10$\%$, and 0.7$\%$, respectively.

\end{abstract}


\pacs{24.10.Lx, 25.75.Dw, 25.75.-q, 24.10.-i}
\keywords{rapidity distribution of protons, UrQMD model, coalescence model, Lorentz transformation}

\maketitle

\section{Introduction and model settings}
Mainly in order to explore the possible (order of) phase transition from the hadron gas (HG) to the quark-gluon plasma (QGP), people are paying more attention to heavy ion collisions (HICs) in the beam energy region from several to several tens GeV$/$nucleon which are currently experimentally covered by BNL Alternating Gradient Synchrotron (AGS), CERN Super Proton Synchrotron (SPS), as well as the Beam Energy Scan (BES) program of BNL Relativistic Heavy Ion Collider (RHIC). Meanwhile, quite a few probes, such as charmonium suppression \cite{Matsui:1986dk}, strangeness enhancement \cite{Soff:1999et}, directed flow \cite{Steinheimer:2014pfa}, elliptic flow (as well as its difference between particles and its anti-partners) \cite{Sorge:1998mk,Torrieri:2007qy,Steinheimer:2012bn,Xu:2013sta},  and Hanbuary-Brown-Twiss (HBT) two-particle correlation \cite{Adamova:2002ff,Li:2007yd,Li:2010ew,Adamczyk:2014mxp}, have been suggested as signals to detect the possible (phase) transition. Among them, the experimental observables related to protons should be theoretically investigated and described well firstly since nucleons are initial particles and heavily influenced by the whole dynamical evolution process and all other newly produced particles come directly or indirectly from the collisions between nucleons. However, it is noticed that even the yields of free (net) protons emitted from HICs at AGS and SPS energies \cite{Feng:2009zzb} are not well described in the framework of microscopic transport models such as the Ultra-relativistic Quantum Molecular Dynamics (UrQMD) model \cite{Yuan:2010ad} (using the cascade mode, and called UrQMD/C), which will be focused in this paper. In the FOPI experimental article of Ref.~\cite{Reisdorf:2010aa}
when checking some global characteristics of central Au+Au collisions as a function of beam energy, it was found that the percentage of protons in clusters is still about one third of the available charge at the beam energy $E_b=1.5$ GeV$/$nucleon. Therefore, the percentage of clustered protons at higher beam energies such AGS and even SPS deserves attention as well.

It is known that, at GSI Schwerionen Synchrotron (SIS) energies, a conventional phase-space coalescence model ~\cite{Kruse:1985pg,Li:2005kqa,Wang:2013wca} is successfully incorporated with transport models (mainly the QMD-like models) after a proper reaction time $t_{\mathrm{cut}}$ in order to describe multiplicities of clusters. In this afterburner the nucleons with relative momenta $\delta p<P_0$ and relative distances $\delta r<R_0$ will be considered to belong to one cluster. Effects of binding energy, isospin, etc., could be taken into account \cite{Neubert:1999sv,Zhang:2012qm} but are ignored in the current work for simplicity. And, baryons other than nucleons could be treated in a similar  way. In the past calculations, the values for the parameter set ($R_0$, $P_0$) might be chosen in the range of (2.8-3.5 fm, 0.25-0.3 GeV$/$c) in order to better reproduce experimental data. Currently, the values should be enlarged slightly due to a much higher excitation energy for clusters from HICS at AGS and SPS energies. It will be found that only one set of parameters, ($3.8$ fm, 0.3 GeV$/$c), can describe the rapidity distribution of free (net) protons from central Au+Au collisions at AGS and Pb+Pb collisions at SPS energies fairly well, with the help of a mean-field potential version of UrQMD (called UrQMD/M, and see Refs.~\cite{Li:2007yd,Li:2010ew,Li:2010ie} for details). In addition, for each reaction, more than 10 thousand events are calculated in the transport program and stopped at $t_{\mathrm{cut}}=50$ fm$/c$.

\begin{figure}[htbp]
\centering
\includegraphics[angle=0,width=0.8\textwidth]{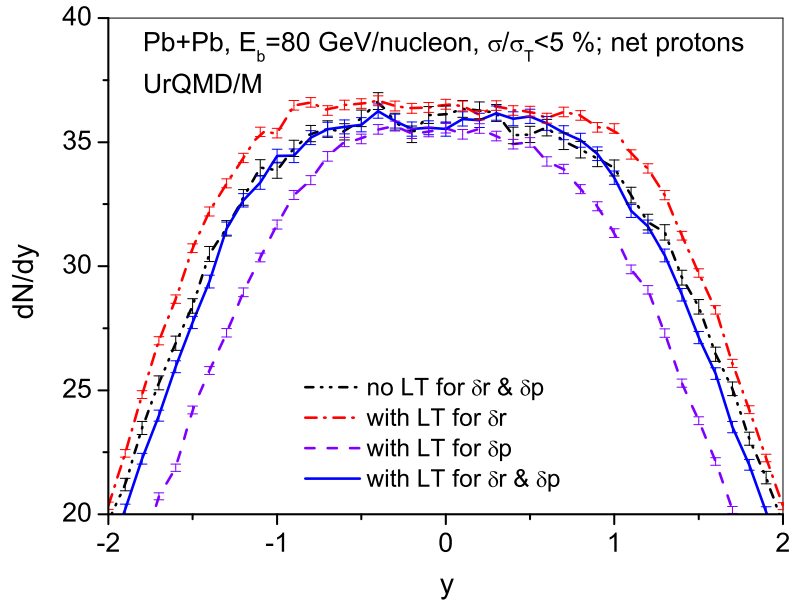}
\caption{\label{fig1} (Color online) Rapidity distribution of net protons from central Pb+Pb collisions at $E_b=80$ GeV$/$nucleon. The $\delta r$ and $\delta p$ in the coalescence afterburner are calculated with or without considering LT (see text for details).
}
\end{figure}

To calculate $\delta r$ and $\delta p$ between two baryons in the coalescence afterburner, the relativistic effect should be taken into account by the well-known Lorentz transformation (LT) from the computational two-nucleus center-of-mass system to the local rest frame of two particles, which is examined in this work for both quantities and shown in Fig.~\ref{fig1}, taking the rapidity $y$ ($=\frac{1}{2}\mathrm{log}(\frac{E_{\mathrm{cm}}+p_{//}}{E_{\mathrm{cm}}-p_{//}})$, where $E_{\mathrm{cm}}$ and $p_{//}$ are the energy and longitudinal momentum of the (anti-)proton in the center-of-mass system, respectively) distribution of net protons ($p-\overline{p}$) from central ($<5\%$ of the total cross section $\sigma_T$) Pb+Pb collisions at $E_b=80$ GeV$/$nucleon as an example. Besides the cases without  (dash-dot-dotted line)  and with (solid line) the consideration of the relativistic effect on both quantities, the effect on each quantity is shown by dash-dotted  (for $\delta r$) and dashed (for $\delta p$) lines, respectively. It is seen clearly that, due to the coordinate-spatial expansion and the momentum-spatial shrinkage by the LT, the net proton yield is visibly enhanced (suppressed) when comparing the dash-dotted (dashed) line to the dash-dot-dotted line, respectively. As a result, the cancellation of the relativistic effect on both quantities is large and makes the final distribution close to that without considering LT for $\delta r$ and $\delta p$ in the afterburner.

\section{Proton yields from UrQMD calculations}
\begin{figure}[htbp]
\centering
\includegraphics[angle=0,width=0.8\textwidth]{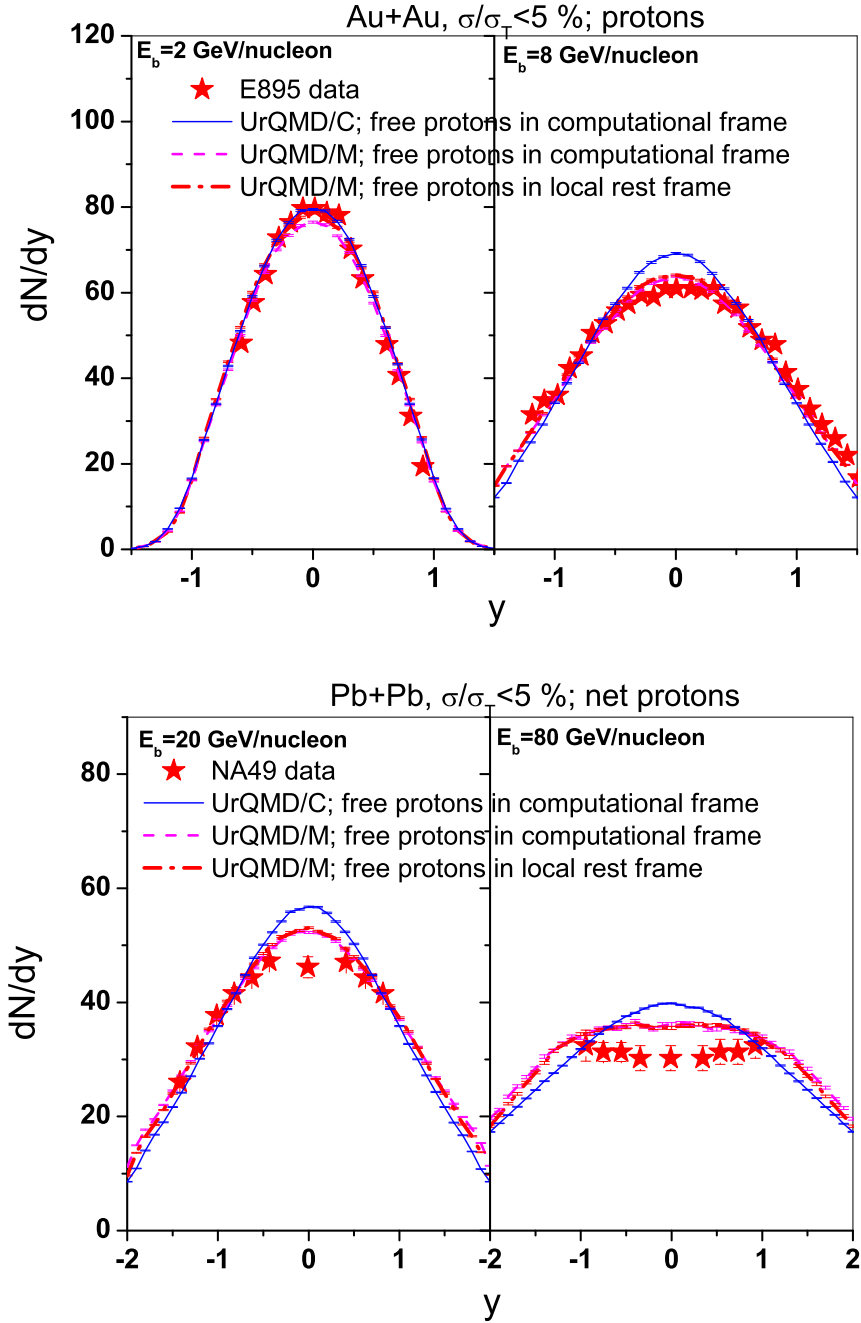}
\caption{\label{fig2} (Color online) Top two plots: Rapidity distribution of protons from central Au+Au reactions at AGS energies 2 and 8 GeV$/$nucleon, respectively. Bottom two plots: Rapidity distribution of net protons from Pb+Pb reactions at SPS energies 20 and 80 GeV$/$nucleon, respectively. Free (net) protons from UrQMD/M after the coalescence burner without and with LT are shown with dashed and dash-dotted lines, while those from UrQMD/C without the LT in the afterburner are shown by solid lines.  The E895 data are taken from Ref.~\cite{Klay:2001tf}. The NA49 data are taken from Refs.~\cite{Blume:2007kw,Strobele:2009nq}.}
\end{figure}

This situation is also true for HICs at other energies. In Fig.~\ref{fig2} we further show the rapidity distribution of protons (top plots) and net-protons (bottom plots) from central ($\sigma/\sigma_T<5\%$) Au+Au reactions at AGS energies 2 and 8 GeV$/$nucleon, and Pb+Pb reactions at 20 and 80 GeV$/$nucleon, respectively. UrQMD calculation results (different lines) are compared with the experimental E895 \cite{Klay:2001tf} and NA49 \cite{Blume:2007kw,Strobele:2009nq} data (stars). First of all, it is seen that the LT in the afterburner modifies the proton yield a little in all plots because of the reason discussed for Fig.~\ref{fig1}. Second, the mean-field potential modifications for both ``pre-formed'' hadrons and formed baryons in UrQMD/M widen the rapidity distributions (as a result, to reduce the yields at mid-rapidity) especially for HICs at higher beam energies, which had been found in the previous calculations \cite{Li:2007yd,Yuan:2010ad}. The additional pressure (and stopping) at the early compression stage leads to, however, less two-body collisions at the later expansion stage and earlier freeze-outs, which results in as a whole weaker stopping power.  Finally, it is interesting to see that calculations with potentials describe both E895 and NA49 data fairly well using only one parameter set of ($R_0$, $P_0$)=($3.8$ fm, 0.3 GeV$/$c) in the coalescence, regardless of the consideration of LT effect, except for those at mid-rapidities and at SPS energies. The discrepancy between UrQMD/M calculations and experimental data at SPS energies leaves a space for a more systematic description of the dynamical evolution of the new phase created at the early stage, such as the stiffness of EoS \cite{Li:2007yd,Li:2008qm,Xu:2013sta} and modifications of cross sections \cite{AbdelWaged:2004wk,Xu:2013sta,Steinheimer:2015sha}.

\begin{figure}[htbp]
\centering
\includegraphics[angle=0,width=0.8\textwidth]{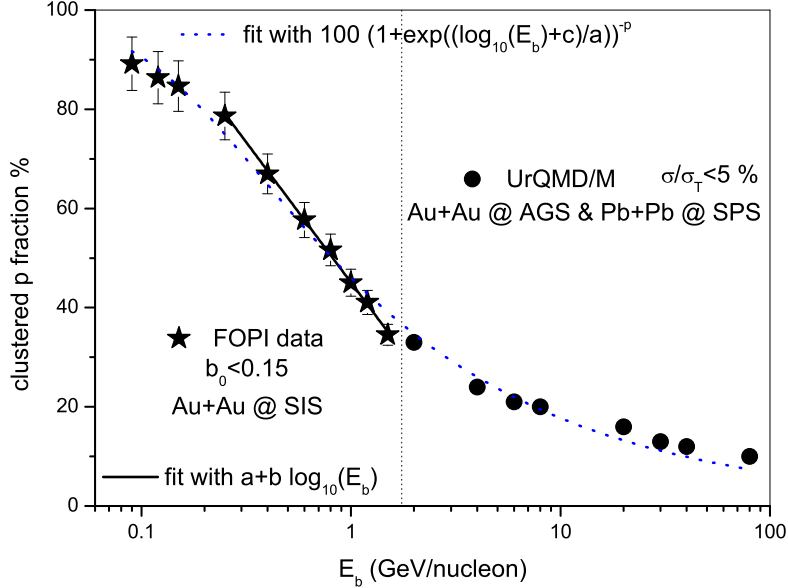}
\caption{\label{fig3} (Color online) Excitation function of the clustered proton fraction (in $\%$). At SIS energies, the FOPI data (stars) are fitted with a function $a+b$log$_{10}$ $(E_{\mathrm b})$ (solid line), while both experimental data at SIS and UrQMD/M calculations at AGS and SPS energies are fitted with a Woods-Saxon-like function (dotted line).
}
\end{figure}

In Ref.~\cite{Reisdorf:2010aa} it was found that, if the proton fraction in clusters to all protons produced from Au+Au collisions at SIS energies and at reduced impact parameters ($b_0<0.15$, where $b_0$ is defined by $b/b_{\mathrm max}$ and $b_{\mathrm max}$ is the sum of both projectile and target sizes) is plotted as a function of beam energy, and the abscissa is set to be logarithmic, the excitation function shows a nicely linear dependence in the energy range from 0.2 to 1.5 GeV$/$nucleon, which is also shown in the left side of Fig.~\ref{fig3} with scattered star symbols. However, if we extrapolate the fitted line (solid) to higher energies, it would be found that there is no clusters any more at the beam energy around 6.5 GeV$/$nucleon, which is obviously not supported by our UrQMD/M calculations (shown in the right side of Fig.~\ref{fig3} with scattered circle symbols). For example, at $E_b=80$ GeV$/$nucleon the proton percentage in clusters keeps still on the order of 10.  Therefore, a Woods-Saxon-like function (dotted line) is used to fit simultaneously both experimental data at SIS and UrQMD/M calculations at AGS and SPS energies, and the fitting result is satisfying with an adjusted coefficient of determination (adj. $R$ square ) of 0.99. It is interesting to see that if we extrapolate the dotted fitting line to RHIC such as the nucleon-nucleon center-of-mass energy $\sqrt{s_{NN}}$=200 GeV (correspondingly, $E_b\simeq 2.1\times 10^4$ GeV$/$nucleon), there are still about 0.7$\%$ of total protons in clusters, which is on the order of the prediction by RQMD calculations with the help of a Wigner function approach \cite{Monreal:1999cp} and measured by STAR and BRAHMS collaborations of RHIC \cite{Abelev:2009ae,Arsene:2010px}.

\section{Summary}
To summarize, with a potential version of the UrQMD in which mean-field potentials for both pre-formed hadrons and confined baryons are considered, and a traditional coalescence model in which one parameter set of ($R_0$,$P_0$)=(3.8 fm, 0.3 GeV$/$c) is used, both E895 proton data at AGS energies and NA49 net proton data at SPS energies can be described reasonably well. And, because of the large cancellation between the expansion in relative distance and the shrinkage in relative momentum through the Lorentz transformation in the coalescence model, the relativistic effect in clusters has little effect on the rapidity distribution of free (net) protons. The calculated excitation function of the proton fraction existing in clusters deviates from a pure logarithmic function as seen by FOPI collaboration at SIS energies. Using a Woods-Saxon-like function, one can fit well both the data at SIS energies and the UrQMD calculation results at AGS and SPS energies. Further, it is found that for central Au+Au or Pb+Pb collisions at top SIS, SPS and RHIC energies, the proton fractions in clusters are about 33$\%$, 10$\%$, and 0.7$\%$, respectively.

\begin{acknowledgements}
We thank Profs. M. Bleicher and Fuqiang Wang for useful discussions and acknowledge support by the computing server C3S2 in Huzhou
University. The work is supported in part by the National
Natural Science Foundation of China (Nos. 11375062, 11275068), the project sponsored by SRF for ROCS, SEM, and the Doctoral Scientific Research Foundation (No. 11447109).
\end{acknowledgements}

\end{document}